\documentstyle[jkas2]{article}

\runningauthor{KANG}
\runningtitle{COSMIC-RAY ACCELERATION} 
\beginpage{1}
\endpage{12}

\def\etal{{\it et al.~}}
\def\eg{{\it e.g.,~}}
\def\ie{{\it i.e.,~}}
\def\kms{~{\rm km~s^{-1}}}
\def\cm3{~{\rm cm^{-3}}}

\def\lsim{\mathrel{  
        \raise0.3ex\hbox{$<$}\kern-0.75em{\lower0.65ex\hbox{$\sim$}}}}
\def\gsim{\mathrel{
        \raise0.3ex\hbox{$>$}\kern-0.75em{\lower0.65ex\hbox{$\sim$}}}}

\begin{document}

\title{ACCELERATION OF COSMIC RAYS AT COSMIC SHOCKS}

\author{Hyesung Kang }

\address{Department of Earth Sciences, 
Pusan National University, Pusan 609-735, Korea}
\address{Department of Astronomy, 
University of Minnesota, Minneapolis, MN 55455, USA \\
{\it E-mail: kang@uju.es.pusan.ac.kr}}
\address{\normalsize{\it (Received March 5, 2003; Accepted March 18, 2003)}}

\abstract{
Nonthermal particles can be produced due to incomplete thermalization 
at collisionless shocks and further accelerated to very
high energies via diffusive shock acceleration.
In a previous study we explored the cosmic ray (CR) acceleration 
at cosmic shocks through numerical simulations of CR modified, 
quasi-parallel
shocks in 1D plane-parallel geometry with the physical parameters relevant for 
the shocks emerging in the large scale structure formation of the universe
(Kang \& Jones 2002).
Specifically we considered pancake shocks driven by 
accretion flows with $u_{\rm o}=1500\kms$ and the preshock gas temperature of 
$T_0=10^4-10^8$K.
In order to consider the CR acceleration at shocks with a broader range
of physical properties, in this contribution we present additional simulations 
with accretion flows with $u_{\rm o}= 75-1500 \kms$ and $T_0=10^4$K.
We also compare the new simulation results with those
reported in the previous study. 
For a given Mach number, shocks with higher speeds accelerate CRs
faster with a greater number of particles,
since the acceleration time scale is $t_{acc} \propto u_{\rm o}^{-2}$. 
However, two shocks with a same Mach number but with different shock
speeds evolve qualitatively similarly when the results are presented in
terms of diffusion length and time scales.
Therefore, the  time asymptotic value for the fraction of shock kinetic energy 
transferred to CRs is mainly controlled by shock Mach number 
rather than shock speed. 
Although the CR acceleration efficiency depends weakly on a well-constrained
injection parameter, $\epsilon$, and on shock speed for low shock Mach 
numbers, the dependence disappears for high shock Mach numbers.  
We present the ``CR energy ratio'', $\Phi(M_s)$, for a wide range
of shock parameters and for $\epsilon=0.2-0.3$ at terminal time of
our simulations. We suggest that these values can be considered as 
time-asymptotic values for the CR acceleration efficiency,
since the time-dependent evolution of CR modified shocks has become
approximately self-similar before the terminal time.  
}

\maketitle

\section{INTRODUCTION}
Collisionless shocks form ubiquitously in astrophysical environments via
collective electromagnetic viscosities,
when supersonic disturbances propagate into tenuous, magnetized, cosmic plasmas.
Due to incomplete plasma thermalization at collisionless shocks,
some suprathermal particles leak upstream and
their streaming motions against the background fluid 
generate strong MHD Alfv\'en waves upstream of the shock 
(e.g., Wentzel 1974; Bell 1978;Lucek \& Bell 2000).
Although these self-excited MHD waves provide necessary electromagnetic 
viscosities to confine thermal particles to the downstream region of the shock, 
some suprathermal particles in the high energy tail of 
the Maxwellian velocity distribution may re-cross the shock upstream.
Then these particles are scattered back downstream 
by those same waves and can be accelerated further to higher energies 
via Fermi first order process.
Here we refer to cosmic rays as nonthermal particles above the Maxwellian
distribution in momentum space, so they include nonrelativistic,
suprathermal particles as well as usual relativistic particles.
It is believed that cosmic ray particles are natural byproducts of the
collisionless shock formation process, and they are extracted from the 
shock-heated thermal particle distribution (Malkov \& V\"olk 1998,
Malkov \& Drury 2001).

According to diffusive shock acceleration (DSA) theory, 
a significant fraction of the kinetic 
energy of the bulk flow associated with a strong shock can be converted 
into CR protons, depending on the CR injection rate
(\eg Drury 1983; Berezhko, Ksenofontov, \& Yelshi 1995; Kang \& Jones 1995). 
In Gieseler, Jones \& Kang (2000), we developed a numerical
scheme that self-consistently incorporates a ``thermal leakage'' injection
model based on the analytic, nonlinear calculations of Malkov (1998).
This injection scheme was then implemented into the combined gas dynamics 
and the CR diffusion-convection code with the Adaptive Mesh Refinement 
technique by Kang, Jones \& Giesler (2002).
In Kang \& Jones (2002) (Paper I, hereafter) we applied this code
to study the cosmic ray acceleration at shocks by numerical simulations 
of CR modified, quasi-parallel shocks in 1D plane-parallel geometry 
with the physical parameters relevant for the cosmic shocks emerging 
in the large scale structure formation of the universe.
We adopted the Bohm diffusion model for CRs, based on the hypothesis that strong
Alfv\'en waves are self-generated by streaming CRs.
We found in Paper I 
that about $10^{-3}$ of incoming thermal particles are injected into the CR
population with our thermal leakage injection model
and up to 60 \% of initial shock kinetic energy is transferred to CRs for 
strong shocks.

During the formation of large scale structure in the universe
cosmic shocks are produced by flow motions
associated with the gravitational collapse of nonlinear structures
(\eg Kang \etal 1994a; Miniati \etal 2000).
These structures are surrounded by accretion shocks 
and CRs can be accelerated to very high energies
at these shocks via the first order Fermi process (Kang, Ryu \& Jones 1996,
Kang, Rachen, \& Biermann 1997, Miniati \etal 2001a, b).
In a recent study, Ryu, Kang, Hallman, \& Jones (2003)
studied the characteristics of cosmic shocks in a cosmological
hydrodynamic simulation of a $\Lambda$CDM universe and their roles on
thermalization of gas and acceleration of nonthermal particles.
They showed that cosmic shocks have a wide range of physical 
parameters with shock speed up to $\sim 3000\kms$, 
the preshock temperature of $10^4< T_0< 10^8$, and 
Mach number up to a few 100.
In this contribution we extend our study in Paper I by including
simulations of shock models with a broader range of physical parameters 
that are relevant for such cosmic shocks.

In the next section we briefly describe our numerical simulations. 
The simulation results are presented and discussed in \S III,
followed by a brief summary in \S IV.


\section{Numerical Simulations}
\subsection{Numerical Methods}
Our numerical code was described in detail in Paper I, so here we
briefly summarize some special features that are different
from standard hydrodynamics codes:  
1) We solve the gasdynamic equations with CR pressure terms
added in the conservative, Eulerian formulation for one dimensional 
plane-parallel geometry along with the diffusion-convection equation
for the CR momentum distribution function (Kang \etal 2001).
2) An adaptive mesh refinement technique is applied to shock fronts
that are tracked as discontinuous jumps by a front-tracking
method (Kang \etal 2001).
3) An additional equation for the ``Modified Entropy'' is solved to 
follow accurately the adiabatic changes outside of shocks, particularly
in the precursor region (Kang \etal 2002).
4) We adopt an injection scheme based on a ``thermal leakage'' model 
that transfers a small proportion of the thermal proton flux 
through the shock into low energy CRs (Gieseler \etal 2000).
There is a free parameter in the adopted injection model:  
$\epsilon = B_0/B_{\perp}$, defined to measure the ratio of the amplitude 
of the postshock MHD wave turbulence $B_{\perp}$ to the general magnetic field 
aligned with the shock normal, $B_0$ (Malkov and V\"olk, 1998). 
This parameter is rather well constrained, since $0.3\lsim \epsilon \lsim 0.4$ 
is indicated for strong shocks (Malkov \& V\"olk 1998).
However, such values lead to very efficient initial injection and 
most of the shock energy is transfered to the CR component for strong shocks of 
$M_s \gsim 30$ (Kang \etal 2002), causing a numerical problem at the very 
early stage of simulations.
So in this study we consider models with $\epsilon=0.2$ and 0.3 and
then discuss how the CR acceleration depends on $\epsilon$.

\subsection{Diffusion Coefficient, Time, and Length Scales}
As in Paper I we assume a Bohm type diffusion model in which
the diffusion coefficient is given as 
\begin{equation}
\kappa(p) = \kappa_{\rm o} {{p^2}\over {(p^2+1)^{1/2}}},
\end{equation}
where $p$ is expressed in units of $m_{\rm p}c$,
$\kappa_{\rm o}= 3.13\times 10^{22} {\rm cm^2s^{-1}} B_{\mu}^{-1}$,
and $B_{\mu}$ is the magnetic field strength in units of microgauss.
This assumption is based on the hypothesis that strong
Alfv\'en waves are self-generated by streaming CRs and provide
random scatterings strong enough to scatter particles within
one gyration radius.
In order to model amplification of self-generated turbulent waves
due to compression of the perpendicular component of the magnetic field,
the spatial dependence of the diffusion is modeled as
$\kappa(x,p) = \kappa(p)\rho_0/ \rho(x)$,
where $\rho_0$ is the upstream gas density.

With the Bohm diffusion model the mean acceleration time scale for 
a particle to reach momentum $p$ is related to the diffusion 
coefficient and shock speed as
\begin{equation}
\tau_{acc} \approx { {8\kappa (p)} \over V_s^2}\approx (7.9\times 10^9 {\rm years})
({p_{10}}) B_{\mu}^{-1} V_{1000}^{-2},
\end{equation}
where $p_{10}= p / 10^{10}$ and $V_{1000} = V_s/1000 \kms$ is shock
speed in units of $1000\kms$.
Although this expression is valid for strong shocks, it gives a
reasonable estimate for any shock strength within a factor of a few (see Paper I).
The so-called diffusion length of CR protons is given by
\begin{equation}
D_{\rm diff}= {\kappa(p) \over V_s}\nonumber\\
\approx (1.0 {\rm Mpc}) {p_{10}} B_{\mu}^{-1} V_{1000}^{-1}. 
\end{equation}
Thus, for a given value of $\kappa_{\rm o}$, the CR acceleration 
proceeds faster and CR particles diffuse on smaller length scales 
at shocks with greater speeds.
Consequently, it is more convenient to normalize the simulation variables in
terms of a diffusion time, $t_{\rm o}=\kappa_{\rm o}/u_{\rm o}^2$, 
and a diffusion length, $x_{\rm o}=\kappa_{\rm o}/u_{\rm o}$ 
(where $u_{\rm o}$ is a characteristic velocity in the problem),
rather than adopting constant values across different models. 
They represent the diffusion length and time scales for the 
protons of $p=1$.
In our discussion a choice of $\kappa_{\rm o}$ (or $B_{\mu}$) is arbitrary,
since we focus on the CR acceleration efficiency at time asymptotic
limits. 

\subsection{1D Plane-parallel Shock Models}
In general, cosmic shocks associated with the large scale structure formation
can be oblique and have various geometries, depending on types of nonlinear
structures onto which accretion flows fall. 
As in Paper I, however, we study the CR acceleration 
at one-dimensional (1D) quasi-parallel shocks which form by 
accretion flows in a plane-parallel geometry in this paper. 
Since our simulations follow the acceleration of CRs up to only 
mildly relativistic energies, the diffusion length scales are much 
smaller than typical curvature radius of cosmic shocks.
So, the diffusion and acceleration of CRs can be studied with
1D plane-parallel shock models.

We consider a pancake shock formed by a steady accretion flow with
a constant density and gas pressure:  
a 1D simulation box with [0,~$x_{\rm max}$] and
an accretion flow entering into the right boundary of the
simulation box with a constant density,
$\rho_0$, gas pressure, $P_{\rm g,0}$, and velocity, $u_0$.
There are no pre-existing CRs in the accretion flows.
The accretion flow Mach number is defined by $M_0 = |u_0|/c_s$, 
where $c_s= (\gamma P_{\rm g,0}/\rho_0)^{1/2}$ is the sound speed of the
accreting gas.
The flow is reflected at the left boundary ($x=0$) and 
a shock propagates to the right. 
For a hydrodynamic shock without the CR pressure, the shock speed
is $u_s = |u_0|/(r-1)$ in the simulation frame
and $u_s^\prime=|u_0|r/(r-1)$ in the far upstream rest frame,
where $r$ is the compression ratio across the shock.
For the accretion Mach number, $2\le M_0\le 100$,
the {\it initial} shock speed ranges $u_{s,0}^\prime=(4/3-3/2)|u_0|$ 
before the CR pressure is built up. 
So the {\it initial} shock Mach number is given by $M_s = (4/3 -3/2)M_0$.  
After we include the CR acceleration, the CR pressure feedback 
slows down the shock, so 
the {\it instantaneous} shock speed, $u_s^\prime(t)$,
in the far upstream rest frame decreases over time.
A CR modified shock consists of a smooth precursor and a subshock,
since CRs diffuse upstream of the subshock, and the CR pressure
decelerates and heats the preshock flow adiabatically,
resulting in weakening of the subshock.

\subsection{Constant $u_0$ {\it versus} Constant $T_0$ Models}
In Paper I we considered a set of models in which the infall speed 
was fixed at $u_0=-1500\kms$, while the preshock temperature was varied
according to $T_0 = 10^4 K (M_0/100)^{-2}$, 
where the accretion flow Mach numbers in the range of  $2\le M_0\le 100$ 
were considered. 
So, in Paper I we calculated CR modified shocks with different 
preshock temperature but with the same accretion speed.
We will call this group of models ``constant $u_0$'' models,
which was designed to study shocks with a wide range of 
the preshock temperature and a canonical velocity inside intracluster
medium of galaxy clusters. 

In this paper we consider accretion flow models with a same preshock 
temperature at $T_0=10^4$ but with different shock speeds according to
\begin{equation}
|u_0|= 1500 \kms (M_0/100),
\end{equation}
which will be referred as ``constant $T_0$'' models.
This set of the constant $T_0$ model is 
designed to understand the CR acceleration at shocks that
have a same preshock temperature but a wide range of shock velocity.
The preshock temperature $T_0=10^4$ is chosen to represent the mean
temperature of the photoionized plasma after reionization of 
intergalactic medium at $z\sim 5$ (Gnedin 2000).
We consider models with $5\le M_0 \le 50$, corresponding to
$75\kms < |u_0|< 750 \kms$.
The model with $M_0=100$ of the constant $T_0$ models is identical to 
that of the constant $u_0$ models presented in Paper I, so not considered here.

\subsection{Normalization of Physical Variables}
\begin{figure*}[t]
\vskip 0.0cm
\centerline{\epsfysize=12cm\epsfbox{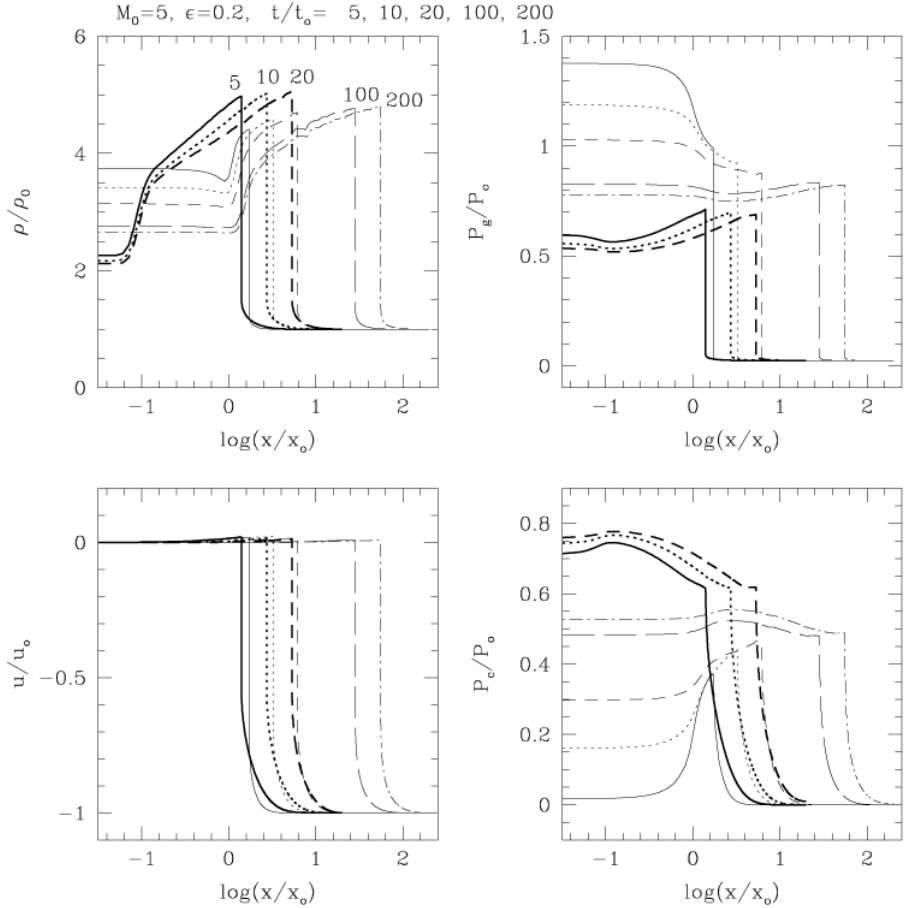}}
\vskip 0.0cm
\caption{
Time evolution of the shocks driven by the accretion flows with $M_0=5$.
Heavy lines represent the model with $T_0=10^4$K and $u_{\rm o}=75\kms$ at
$t/t_{\rm o}=5,$ 10, and 20. 
Light lines show the model from Paper I with $T_0=4\times 10^6$K and
$u_{\rm o}=1500\kms$ at $t/t_{\rm o}=5,$ 10, 20, 100, and 200. 
The diffusion time scale, $t_{\rm o}=\kappa_{\rm o}/u_{\rm o}^2$, 
and the diffusion length, $x_{\rm o}=\kappa_{\rm o}/u_{\rm o}$ are 
defined by the accretion speed of each model, $u_{\rm o}$.
The inverse wave-amplitude parameter $\epsilon=0.2$ is adopted for
both models.
The accretion flows are reflected at $x/x_{\rm o}=0$ and 
the shocks of $M_s=6.8$ propagate to the right, 
so the leftmost profile corresponds to the earliest time.
Note the distance from the reflecting plane is in a logarithmic scale. 
}
\end{figure*}
\begin{figure*}[t]
\vskip 0.0cm
\centerline{\epsfysize=12cm\epsfbox{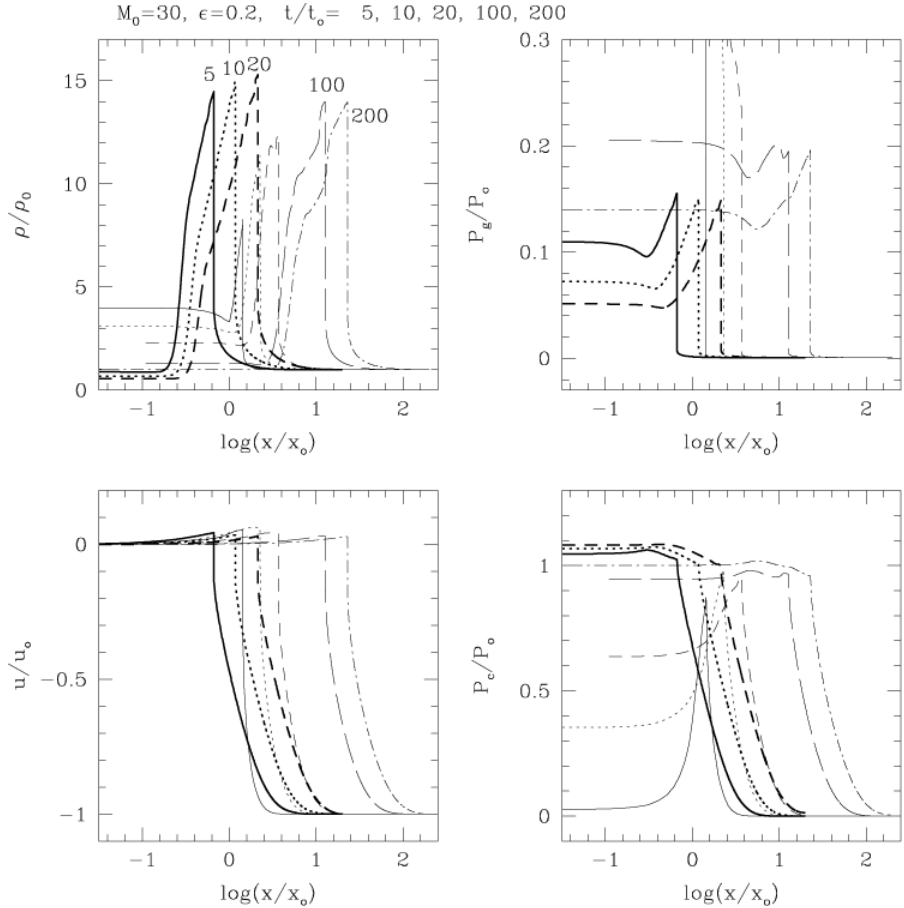}}
\vskip 0.0 cm
\caption{
Same as Figure 1 except $M_0=30$.
Heavy lines represent the model with $T_0=10^4$K and $u_{\rm o}=450\kms$ at
$t/t_{\rm o}=5,$ 10, and 20.
Light lines represent the model  with $T_0=4.4\times 10^5$K and
$u_{\rm o}=1500\kms$ at $t/t_{\rm o}=5,$ 10, 20, 100, and 200.
}
\end{figure*}
\begin{figure*}[t]
\vskip -0.5cm
\centerline{\epsfysize=14cm\epsfbox{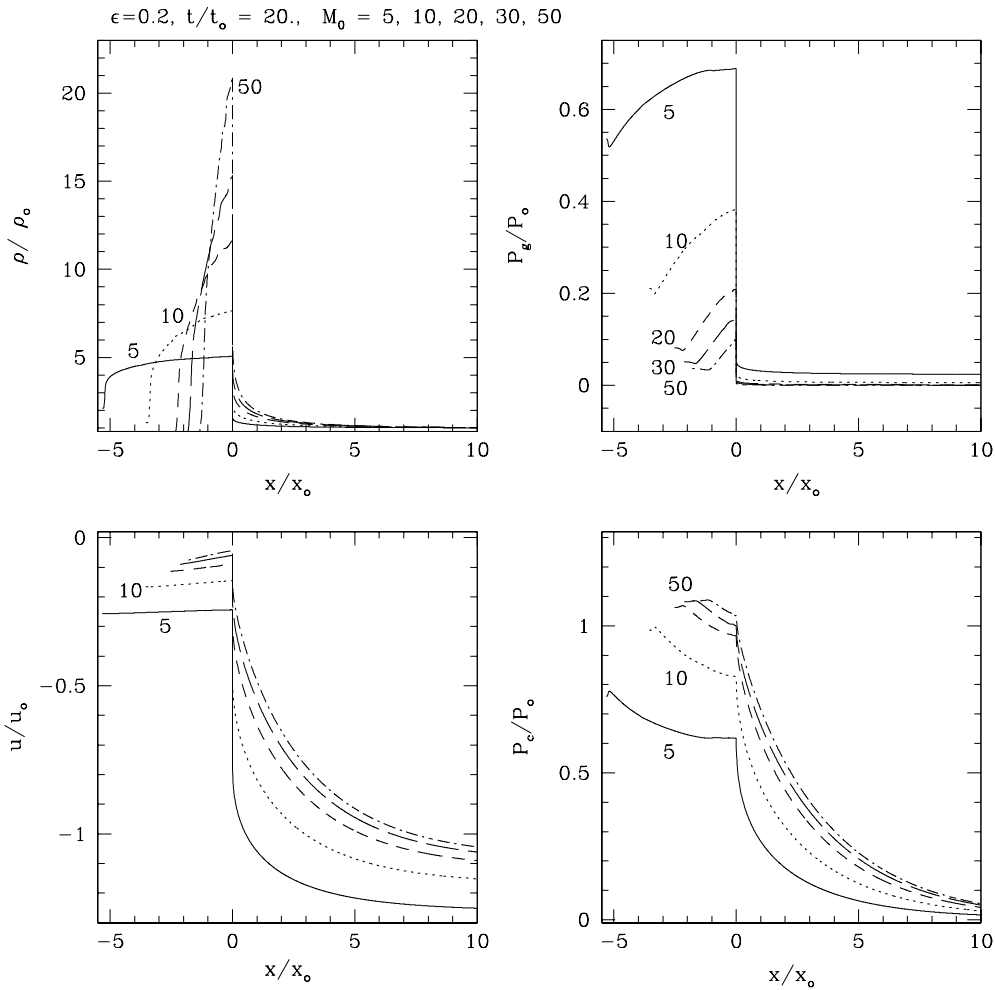}}
\vskip -0.5 cm
\caption{
Structure of cosmic-ray modified shocks driven by the accretion flows 
with $T_0=10^4$K at $t/t_{\rm o}=20$.
The accretion flow speed scales with the accretion Mach number as
$u_0=1500\kms (M_0/100)$.
The inverse wave-amplitude parameter $\epsilon=0.2$ is adopted. 
The length and time coordinates are expressed in units
of $x_{\rm o}$ and  $t_{\rm o}$, respectively, 
defined by the accretion speed of each model. 
}
\end{figure*}
\begin{figure*}[t]
\vskip -0.5cm
\centerline{\epsfysize=14cm\epsfbox{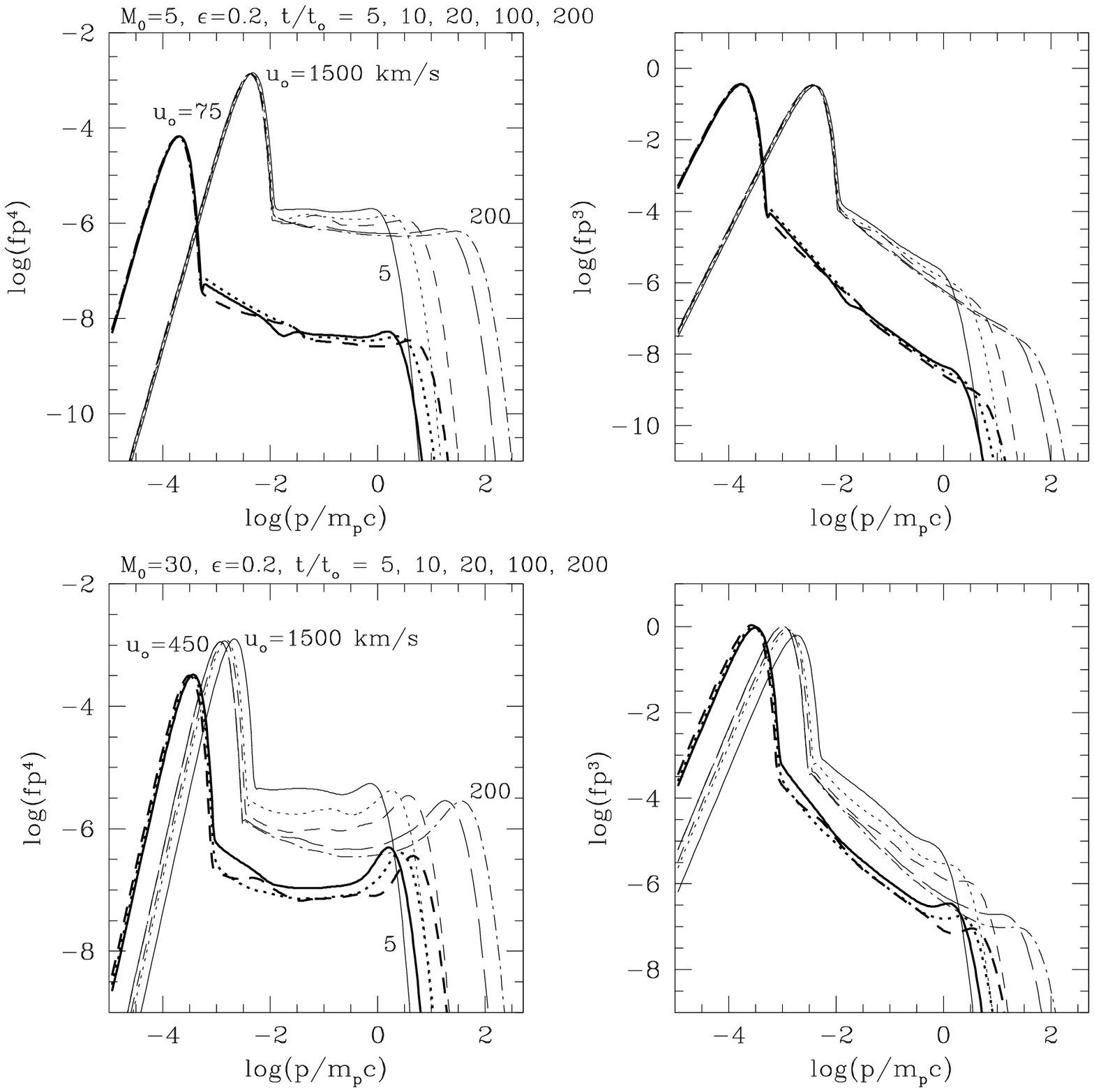}}
\vskip -0.5cm
\caption{
Evolution of the particle distribution function {\it at the shock}, 
represented as $p^4f(p)$ and $p^3f(p)$, is shown 
for two sets of the models shown in Figures 1-2. 
Heavy lines represent the shock models with $T_0=10^4$K at 
$t/t_{\rm o}=5,$ 10, and 20. The accretion flow speed is
$u_{\rm o}=75\kms$ for $M_0=5$ and $u_{\rm o}=450\kms$ for $M_0=30$.
Light lines represent the shock models with $u_{\rm o}=1500\kms$ 
at $t/t_{\rm o}=5,$ 10, 20, 100, and 200.
The peaked distribution at low momenta ($p<10^{-2}$) represent
the thermal Maxwellian distribution.
The same line types are used here as in Figures 1-2.
}
\end{figure*}

Ideal gasdynamic equations in 1D planar geometry do not contain any intrinsic
time and length scales, but in CR modified shocks the CR acceleration and 
the precursor growth develops over the diffusion time on diffusion
length scales.  
So, there are intrinsic similarities in the dynamic evolution and structure
of two CR shock models with a same Mach number but with different preshock
temperature or with different shock speeds, 
if we present the results in the 
coordinate system normalized with diffusion scales 
($t_{\rm o}$ and $x_{\rm o}$). 

As in Paper I we set the far upstream density and flow values as 
$\tilde \rho_0=\rho_0/\rho_{\rm o}=1$, $\tilde u_0=u_0/u_{\rm o}= -1$ 
in code units for all models, where $\rho_{\rm o}$ and $u_{\rm o}$
are normalization constants for the gas density and the flow speed, 
respectively. 
The normalization constants depend on the accretion Mach number $M_0$ 
as follows: $u_{\rm o}=1500~\kms (M_0/100)$ and 
so $\beta=u_{\rm o}/c = 0.005 (M_0/100)$.
Thus the length and time scales depend on $M_0$ according to 
$x_{\rm o}=\kappa_{\rm o}/u_{\rm o}\propto M_0^{-1} $ and
$t_{\rm o}=\kappa_{\rm o}/u_{\rm o}^2 \propto M_0^{-2}$, respectively, 
with an arbitrary choice of diffusion coefficient, $\kappa_{\rm o}$. 
The gas density normalization constant, $\rho_{\rm o}$, is arbitrary 
as well, but the pressure normalization constant 
depends on $M_0$ as 
$P_{\rm o}=\rho_{\rm o} u_{\rm o}^2 \propto M_0^2$. 
Throughout the paper and in the code  
physical variables are given in units of the normalization
constants, $x_{\rm o}$, $t_{\rm o}$, $u_{\rm o}$, $\rho_{\rm o}$,
and $P_{\rm o}$.

The models with smaller $M_0$ have smaller shock speeds and lower postshock
temperature, so typical shock thickness, 
which is of order of gyro radius of the thermal protons,
is smaller than that of the models with larger $M_0$. 
Since an effective numerical shock thickness is the grid spacing at the
finest grid level, for low $M_0$ models, 
the required grid spacing in current simulations 
is much smaller than that of Paper I. 
The new simulations are carried out on a base grid with
$\Delta x_0 = 0.002$ using $l_{\rm max}=7$ additional grid levels,
so $\Delta x_7 = 1.56\times 10^{-5}$ at the finest grid level.
This leads to a severe requirement of computation time, so
we run the simulations for $t/t_{\rm o}=20$, much shorter time than that
of Paper I, $t/t_{\rm o}=200$.
The simulated space is $x=[0,20]$ and $N=10000$ zones are used
on the base grid. 
The number of refined zones around the shock is $N_{rf}=50$ on the base
grid and so there are $2N_{rf}=100$ zones on each refined level.
To avoid technical difficulties,
the multi-level grids are used only after the shock propagates away from
the left boundary at the distance of $ x_s = 0.05$.
After the shock moves to $x_s=0.05$ (at $t \approx 0.15 $ for strong shocks),
the AMR technique is turned on and the CR injection and acceleration are
activated.
This initial delay of the CR injection and acceleration should not
affect the final outcomes. 
For all models we use 230 uniformly spaced logarithmic momentum zones
in the interval $\log (p/m_p c)=[\log p_0,\log p_1]=[-3.0,+3.0]$
As in our previous studies, the function $g(p)=p^4f(p)$ is evolved instead
of $f(p)$ and $y = \ln p $ is used instead of the momentum variable, $p$
for that step.

\section{RESULTS}
Since the CR acceleration is mainly determined by velocity jump
across the shock (\ie $\Delta p/p \propto \Delta u/c$), which 
is a function of shock Mach number only,
we expect {\it two models with the same Mach number but with different
$u_{\rm o}$} may have qualitatively similar results,
when expressed in terms of diffusion scales. 
The main difference between the two models is the ``effective ''
injection momentum, $p_{\rm inj}/m_p c \sim 0.01 (u_{\rm o}/1500\kms)$.
Here the ``effective '' injection momentum refers to 
a mean value of the momentum range 
over which thermal leakage takes place (Gieseler \etal 2000). 
The diffusion coefficient of injected, nonrelativistic 
CRs, $\kappa(p) \propto p_{\rm inj}^2$ for $p_{inj}\ll 1$,
has different momentum dependence from that of relativistic CRs, 
$\kappa(p) \propto p$ for $p>1$.
So the similarity between the two models could be broken at early
evolutionary stages when the CR pressure is still dominated by 
nonrelativistic particles. 
We attempt to make comparisons of such two models in this section.

\subsection{Modified Shock Structure}
We show the time evolution of the shock driven by the accretion flow
with $M_0=5$ and $u_{\rm o}=75\kms$ at $t/t_{\rm o}=5,$ 10, 20 in 
Figure 1 (heavy lines).
The model with $M_0=5$ and $u_{\rm o}=1500\kms$ from Paper I is also
shown at $t/t_{\rm o}=5,$ 10, 20, 100 and 200 (light lines).
As discussed in the previous section, the results are presented in two 
different sets of physical coordinates, 
so, for example, the ratio of diffusion length scales between two model
is $x_{\rm o}(75\kms)/x_{\rm o}(1500\kms)=20$. 
Since we integrate the model with $u_{\rm o}=75\kms$ up to
$t/t_{\rm o}=20$ when nonrelativistic particles still contribute 
significantly to the CR pressure, 
we expect to see some differences in the two models before that time.
First of all, in the model with $u_{\rm o}=75\kms$, 
CR particles are injected at much lower injection momenta 
and their acceleration time scales in units of $t_{\rm o}$ are shorter. 
So the CR pressure increases faster and the modification to
the flow structure due the CR pressure occurs earlier, compared
to the model with $u_{\rm o}=1500\kms$. 
In both models a substantial precursor develops upstream of the shock,
resulting in a significant modification of flow structure.
Also, the shock structures seem to have reached approximate time-asymptotic 
states by the terminal time of both simulations ($t/t_{\rm o}=20$ for
$u_{\rm o}=75\kms$ model and $t/t_{\rm o}=200$ for $u_{\rm o}=1500\kms$ model).
Once the postshock CR pressure becomes constant, the shock structure 
evolves approximately in a ``self-similar'' way, because the scale 
length of shock broadening increases linearly with time.
By comparing the shock structures at the terminal times, we 
expect that the overall evolution and shock structure would approach to
time asymptotic states that are similar for the two models.

In Figure 2 we make the similar comparison between the two models
with $u_{\rm o}=450\kms$ and $u_{\rm o}=1500\kms$ for strong
shocks driven by the accretion flow with $M_0=30$. 
Due to stronger nonlinear feedback effects, 
the overall shock structure displays greater differences 
at a given value of $t/t_{\rm o}$, 
compared to the models with $M_0=5$.
For example, distances of the shock from the reflecting plane are 
different and the gas and CR pressure have very different profiles
before $t/t_{\rm o}\le 20$.
However, the time asymptotic values of the CR pressure become similar
at the terminal times of the two models. 

Figure 3 shows the shock structure of the constant $T_0$ models
with different $M_0$ at the terminal time ($t/t_{\rm o}=20$).
Here the spatial coordinate is shifted so the shock is located at
$x/x_{\rm o}=0.0$ for all models. 
The degree of nonlinear modification and density enhancement
are greater in models with higher $M_0$ due to the higher CR pressure, 
so the distance of the shock from the reflecting plane (the leftmost
point) is smaller.
The CR pressure seems to converge to a limiting
value of $P_c/\rho_{\rm o} (u_{s,0}^\prime)^2\sim 0.6$,
where $u_{s,0}^\prime$ is the initial shock speed before the CR
modification slows down the shock.
The effects of nonlinear feedback such as the preshock
deceleration and the compression through the total shock transition can
be compared most clearly in the flow velocity plot.

\subsection{Particle Distribution Function}

\begin{figure*}[t]
\vskip -0.5cm
\centerline{\epsfysize=14cm\epsfbox{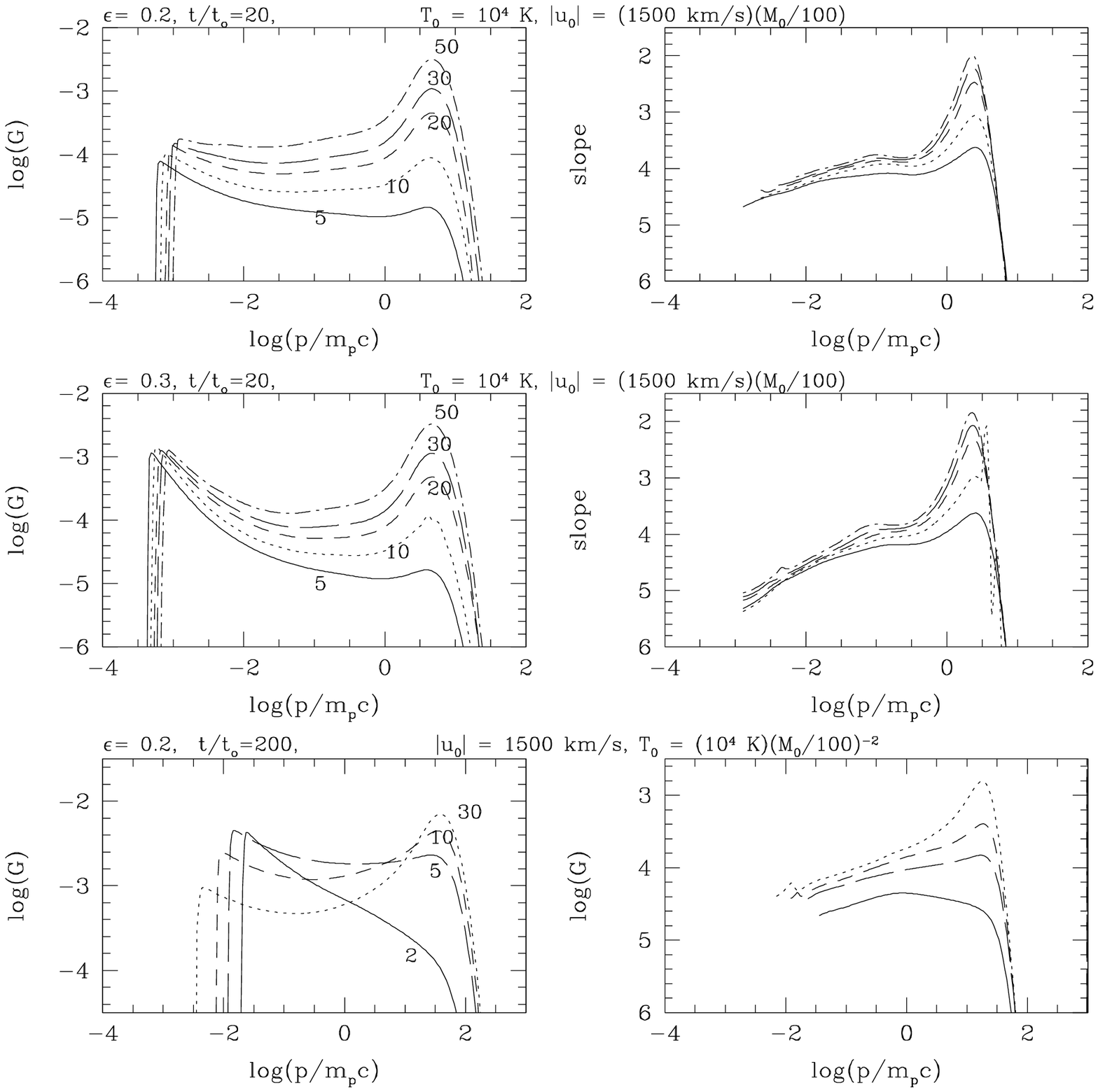}}
\vskip -0.5cm
\caption{
CR distribution function integrated over the simulation
box, $G(p)= \int p^4f_{\rm cr}(p){\rm dx}$, and its power law slope,
$q= -(\partial \ln G / \partial \ln p-4)$, at the terminal time of 
each simulation are shown. 
Top two panels are for the models with $T_0=10^4$ K and $\epsilon=0.2$,
while middle two panels are for the models with $T_0=10^4$ K and $\epsilon=0.3$.
For both sets of models 
the accretion speed is given as $u_{\rm o}=1500\kms (M_0/100)$ and
the terminal time is $t/t_{\rm o}=20$. 
Bottom two panels show the models with $u_{\rm o}=1500\kms$ and $\epsilon=0.2$.
The preshock temperature is given by $T_0=10^4$K$(M_0/100)^{-2}$.
The curves for $G(p)$are labeled with the accretion Mach number.
}
\end{figure*}
\begin{figure*}[t]
\vskip 0.0cm
\centerline{\epsfysize=12cm\epsfbox{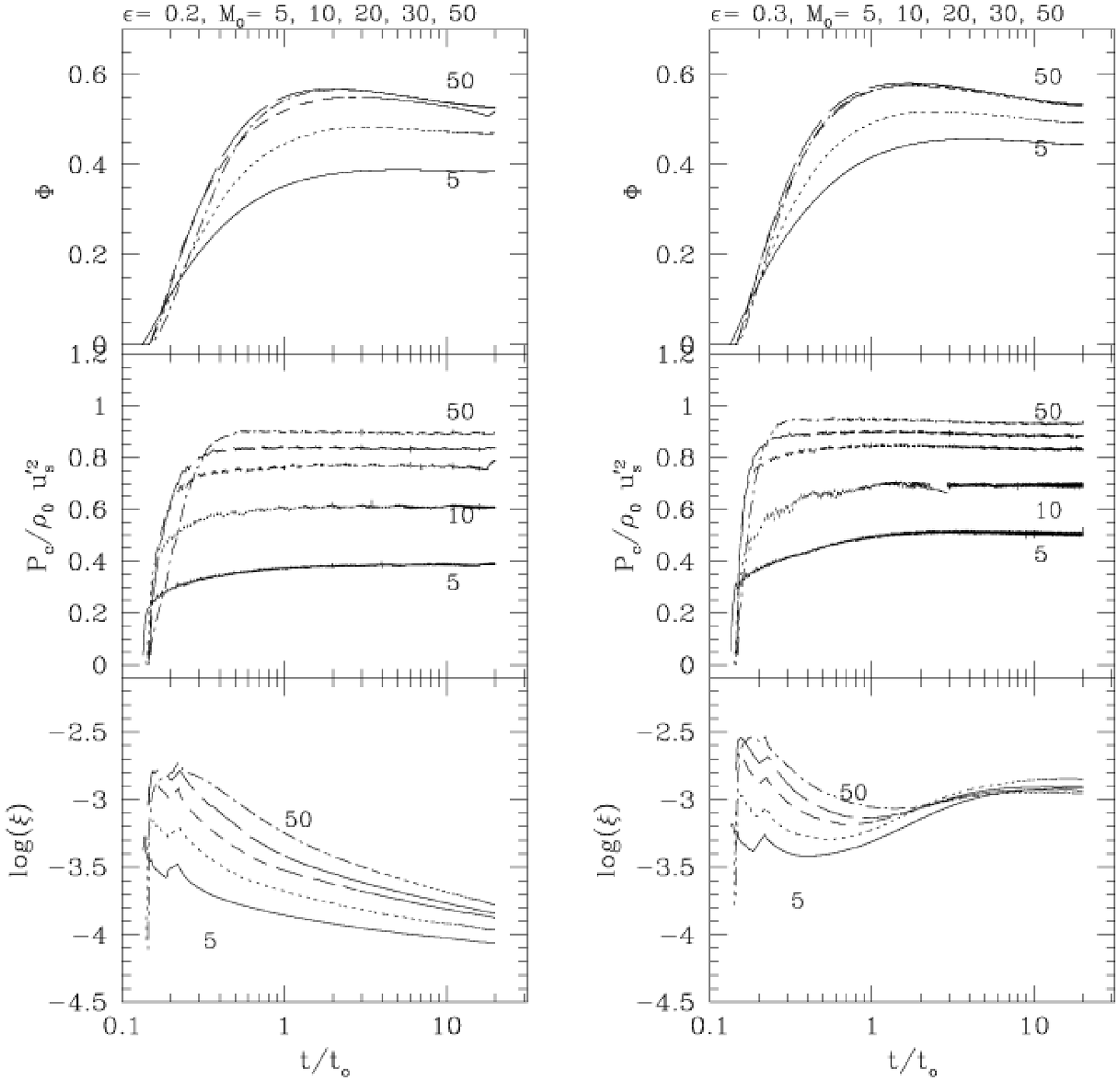}}
\vskip 0.0cm
\caption{
The ratio of total CR energy in the simulation box
to the kinetic energy in the initial shock frame
that has entered the simulation box from upstream, $\Phi(t)$,
the postshock CR pressure in units of far upstream ram pressure in the 
instantaneous shock frame, 
and time-averaged injection efficiency, $\xi(t)$.
Left three panels are for $M_0=5-50$ and  $\epsilon=0.2$.
Right three panels show the same quantities for $M_0=5-50$ and  
$\epsilon=0.3$.
}
\end{figure*}

Figure 4 shows the evolution of the particle distribution functions
(including both thermal and nonthermal populations) 
{\it at the shock} (\ie $f[x_s,p]$)
for two sets of models ($M_0=5$ and $M_0=30$) shown in Figures 1-2. 
In addition to the usual form of distribution function, $g(p)=f(p)p^4$,
we also plotted  $f(p)p^3$ to show the differential number density of
CRs in logarithm momentum bin, $dN = f(p)p^3 {\rm d} \ln p$.
In fact $g(p)$ represent the differential energy density of CRs 
in logarithm momentum bin, $dE = f(p)p^4 {\rm d} \ln p$ for 
relativistic momenta ($p\gg1$).
First we note that peak momentum of thermal distribution is lower 
for the lower postshock temperature, 
so the particle injection starts at lower injection momenta in  
the models with lower $u_{\rm o}$.
For the model with $M_0=30$ and $u_{\rm o}=1500\kms$, we can see that
the Maxwellian distribution shifts to lower momenta
as the postshock temperature diminishes due to energy transfer to CRs. 
For other models this adjustment happens well before $t/t_{\rm o}=5$, so
the postshock thermal distribution is almost steady after that time.

We identify $p_{\rm max}$ as the momentum above which $g(p)$ drops
sharply, characterizing the effective upper cutoff in the CR
distribution. This momentum is approximately related to the age of the shock
in units of the diffusion time as $p_{\rm max} \sim 4.5 (t/t_{\rm o})$
for $p_{\rm max}>1$. 
This explains why values of $p_{\rm max}$ are 
similar at a given value of $t/t_{\rm o}$ for all models, 
regardless of $u_{\rm o}$, injection momentum, or shock Mach number.
In comparison of the two models with a same $M_0$, the one with lower
$u_{\rm o}$ injects CRs at lower momenta, so, for example, 
$\log p_{\rm inj} \approx -3.3$ for $u_{\rm o}=75\kms$, while
$\log p_{\rm inj} \approx -2$ for $u_{\rm o}=1500\kms$. 
Just above the injection pool, the distribution function changes smoothly
from the thermal distribution to an approximate power-law whose index is
close to the test-particle slope for the subshock, 
\ie $q_s = 3r_s/(r_s-1)$ (where $r_s$ is the compression ratio across
the subshock).
The distribution function $g(p)$ shows the
characteristic ``concave upwards'' curves reflecting
modified shock structure (including the precursor) for all models. 
Although the distribution function at a given momentum is larger 
in the higher $u_{\rm o}$ models, but the ratio of
$P_c/\rho _{\rm o} u_{\rm o}^2$ is larger in lower $u_{\rm o}$ models.
This leads to slightly greater nonlinear feedback in lower
$u_{\rm o}$ model, resulting in more concave curves.

Figure 5 shows
the total CR distribution within the simulation box,
$G(p)= \int p^4f_{\rm cr}(p){\rm dx}$ and its power law slope
$q= -(\partial \ln G / \partial \ln p-4)$ {\it at the terminal time} of 
three sets of simulations.
The models shown in top two panels are the constant $T_0$ models with
$\epsilon=0.2$ for $M_0=5,$ 10, 20, 30, and 50.
The particle momenta near $\log p \approx -3.3$ are injection momenta
above the thermal distribution and have similar values within 
a factor of two for models with different $M_0$,
indicating similar thermal populations.
Although the postshock temperature increases with shock speed 
as $T_2 \propto (u_s\prime)^2$ in pure gasdynamic shocks, 
all models in this group have similar $T_2$.
This is because the postshock gas pressure falls and the postshock
density increases due to nonlinear feedback to a greater extent 
at higher Mach number models. 
The integrated distributions also show the characteristic ``concave upwards'' 
curves and this ``flattening'' trend is stronger for higher $M_0$ models.
The slope of the total CR spectrum 
ranges over $4.0 \lsim q \lsim 4.7$ near $p_{\rm inj}$
then decreases with the particle momentum and converges 
for strong shocks to $q \sim 2$ just below $p_{\rm max}$.
The same kind of models but with $\epsilon=0.3$ are shown in middle two
panels.
They show that the injection takes place at slightly lower momenta
with greater numbers of particles in the Maxwellian tail for the models
with larger $\epsilon$, leading to the slope near 
$p_{\rm inj}$ as large as $q \sim 5$.
So the CR distribution functions for $\epsilon=0.3$ models 
are greater than those for $\epsilon=0.2$ models for $p<1$,
while they are roughly similar for $p>1$. 
One can expect that the CR pressure of these two sets of models
may be different at early stage when nonrelativistic, fresh injected
particles dominate the pressure, but it becomes similar once 
relativistic particles dominate.
As in Figure 4, all models shown here have similar values 
of $p_{\rm max}$ regardless of values of $u_{\rm o}$, since the results are 
shown at the same values of $t/t_{\rm o}$.
The bottom two panels show the constant $u_{\rm o}$ models from Paper I.
Since $u_{\rm o}=1500\kms$ for all models, the postshock temperature
would be similar if these are pure gasdynamic shocks.
Due to their higher degree of nonlinear modification to the structure,
however, the postshock thermal populations have lower temperature
in the models with higher $M_0$. 

Although the evolution of these shock becomes approximately self-similar
and the postshock CR pressure reaches a quasi-steady value before
the terminal time, CR particles continue to be accelerated to ever higher
energies and so the CR distribution continues to extend to higher momenta.
Thus the CR distribution functions in Figure 5 show only a snap shot 
at the terminal time. But they illustrate how the CR distribution deviates
from the test-particle like power-law due to nonlinear feedback.

\subsection{Injection and Acceleration Efficiencies}

\begin{figure*}[t]
\vskip -1.0cm
\centerline{\epsfysize=14cm\epsfbox{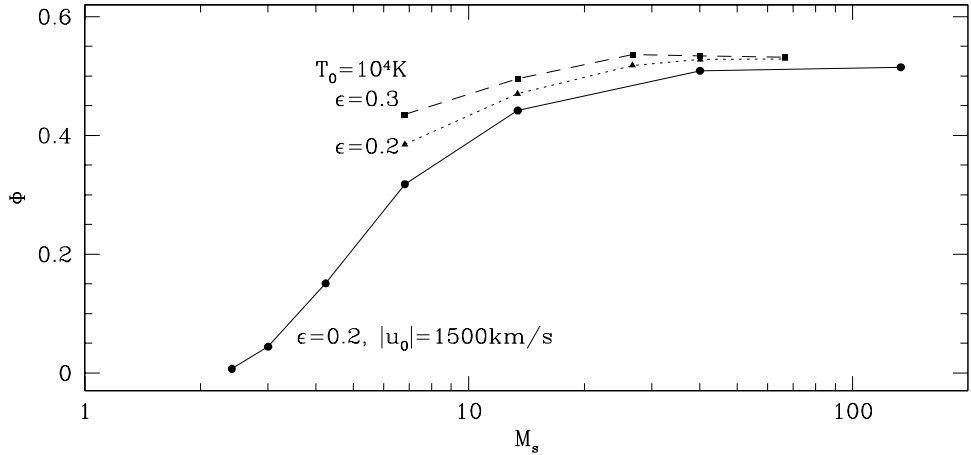}}
\vskip -6.5cm
\caption{
The CR energy ratio, $\Phi$, at the terminal time of the
three sets of simulations shown in Figure 5 as a function 
of the shock Mach number, $M_s$.
The solid line with circles are for models with
$u_{\rm o}=1500 \kms$ and $\epsilon=0.2$, the dotted line with
triangles for models with $T_0=10^4$K and $\epsilon=0.2$,
and the dashed line with squares for models
with $T_0=10^4$K and $\epsilon=0.3$.
}
\end{figure*}

As in Paper I we define the injection efficiency 
as the fraction of particles that have entered the shock from 
far upstream and then are injected into the CR distribution: 
\begin{equation}
\xi(t)=\frac {\int_0^{x_{\rm max}} {\rm d x} \int_{p_0}^{p_1} 4\pi f_{\rm CR}(p,x,t)p^2 {\rm d p}}
{ \int_{t_1}^t n_0 u_s^\prime (t^{\prime}) {\rm d t^{\prime}} }\, 
\end{equation}
where $f_{\rm CR}$ is the CR distribution function, 
$n_0$ is the particle number density far upstream,
$ u_s^\prime$ is the instantaneous shock speed, 
and $t_1$ is the time when the CR injection/acceleration is turned on.

As a measure of acceleration efficiency,
we define the  ``CR energy ratio''; namely the
ratio of the total CR energy within the simulation box to
the kinetic energy in the {\it initial shock frame}
that has entered the simulation box from far upstream,
\begin{equation}
\Phi(t)=\frac {\int_0^{x_{\rm max}} {\rm d x} E_{\rm CR}(x,t)}
 {  0.5\rho_0 (u_{s,0}^{\prime})^3  t },
\label{crenrat}
\end{equation}
where $ u_{s,0}^\prime$ is the initial shock speed. 
Since the shock slows down due to nonlinear modification,
the kinetic energy flux in the instantaneous
shock rest frame also decreases.
So the ratio of the total CR energy in the simulation box
to the kinetic energy defined in the {\it instantaneous
shock frame} which has entered the shock from far upstream, \ie
\begin{equation}
\Phi^\prime (t)= \frac {\int_0^{x_{\rm max}} {\rm d x} E_{\rm CR}(x,t)}
{\int 0.5\rho_0 u_0^\prime (t)^3 {\rm dt}}, 
\end{equation}
can be much larger than $\Phi$, especially for high Mach number shocks.

Figure 6 shows the CR energy ratio, $\Phi$, the CR pressure {\it at the shock} 
normalized to the ramp pressure of the upstream
flow in the instantaneous shock frame,
$P_{\rm c,2}/\rho_0 (u_s^{\prime})^2$,
and the ``time-averaged'' injection efficiencies, $\xi$,
for models with different $M_0$ when $\epsilon=0.2$ 
(left three panels) and $\epsilon=0.3$ (right three panels).
For all Mach numbers the postshock $P_{\rm c,2}$ increases until
a balance between injection/acceleration and advection/diffusion of CRs
is achieved, and then stays at a steady value afterwards.
The time-asymptotic value of the CR pressure becomes
$P_{\rm c,2}/\rho_0 (u_{s,0}^\prime)^2 \sim 0.6$ in the initial shock frame and 
$P_{\rm c,2}/\rho_0 (u_s^\prime)^2 \sim 0.9$ in the instantaneous shock frame 
for $M_0= 50$ with $\epsilon=0.2$.
After $P_{\rm c,2}$ has reached a quasi-steady value and
the evolution of the $P_{\rm c}$ spatial distribution has become
``self-similar'', 
the CR energy ratio also  asymptotes to a constant value. 
Time-asymptotic value of $\Phi$ increases with $M_0$,
but it converges to $\Phi\approx 0.53$ for $ M_0\ge 20$ and $\epsilon=0.2$.
As discussed in detail in Kang \etal (2002), 
the injection rate is higher for higher subshock Mach number 
and for larger values of $\epsilon$, which leads
to the higher CR pressure and higher $\Phi$.
Once again, however, this dependence on $\epsilon$ becomes 
weaker for stronger shocks.
The average injection rate is about $\xi \approx 10^{-4.1} 
-10^{-3.8}$ with $\epsilon = 0.2 $ and 
$\xi \approx 10^{-3}$ with $\epsilon = 0.3$. 

Figure 7 shows values of the CR energy ratio $\Phi$ at the
terminal time of three sets of simulations shown in Figure 5.
Although we were able to follow CR acceleration only up to 
$p_{\rm max}\sim 10-100$, these values can serve as estimates
for the time-asymptotic CR acceleration efficiency,
since $\Phi$ seems to approach constant values after $t/t_{\rm o}>1$.
As shown in Figure 6 the CR acceleration is more efficient
for larger $\epsilon$, so, for the constant $T_0$ models with $M_s=6.8$, 
$\Phi \approx 0.39$ for $\epsilon=0.2$ (dotted line), 
while $\Phi \approx 0.44$ for $\epsilon=0.3$ (dashed line). 
The difference becomes smaller for high Mach number shocks ($M_s\lsim30$).
The constant $T_0$ models have smaller $u_{\rm o}=1500\kms (100/M_0)$ 
for $M_0<100$ than the constant $u_{\rm o}=1500\kms$ models. 
Due to this velocity difference, 
for given values of Mach number and $\epsilon$, 
the constant $T_0$ models have slightly higher $\Phi$ than 
the constant $u_0$ models.
For the constant $u_0$ model with $M_s=6.8$, for example,
$\Phi \approx 0.32$ for $\epsilon=0.2$ (solid line). 
 
\section{SUMMARY}
We have calculated the CR acceleration at 1D quasi-parallel shocks
by using our cosmic-ray AMR Shock code (Kang \etal 2002), 
which incorporates the ``thermal leakage'' injection process to the CR/hydro
code that solves the CR diffusion-convection equation along with CR modified 
gasdynamic equations.
Our simulations have been performed in a 1D plane-parallel space in which
shocks are driven by the accretion flows with 
$u_{\rm o}=75-1500 {\rm km s^{-1}}$ and the temperature of $T_0=10^4$K.
Mach number of the resulting shocks ranges $6.8\le M_s\le 133$. 
We have compared the new simulation results with those from the previous study
in which the accretion flows with
$u_{\rm o}=1500 {\rm km s^{-1}}$ and the temperature of $T_0=10^4-10^{8}$K
were considered (Kang \& Jones 2002, Paper I). 

Detailed simulation results found in Paper I remain valid, so we briefly
review the main conclusions here to make the present paper 
self-contained. 

1) Suprathermal particles can be injected very efficiently
into the CR population via the thermal leakage process, so that typically
a fraction of $10^{-4} - 10^{-3}$ of the particles passed through the shock 
become CRs for $\epsilon = 0.2- 0.3$. 

2) For a given value of shock Mach number, 
the injection efficiency and the CR acceleration are higher 
for larger $\epsilon$. 
But this dependency is weaker for higher Mach numbers of $M_s \gsim 30$. 

3) For a given value of $\epsilon$, the acceleration efficiency
increases with $M_s$, but it asymptotes to a limiting value 
of the CR energy ratio, $\Phi\approx 0.5$, for $M_s \gsim 30$ 
and $\epsilon=0.2-0.3$.

4) The CR pressure seems to approach a steady-state value 
in a time scale comparable to the acceleration time scales
for the mildly relativistic protons after which
the evolution of CR modified shocks becomes approximately ``self-similar''. 
This feature enables us to predict time asymptotic values of the CR 
acceleration efficiency through numerical simulations of CR shock models
with a broad range of the physical parameters in this work.

5) We suggested that the CR acceleration is innate to collisionless 
shock formation process and CRs can absorb a significant fraction of initial 
shock kinetic energy.

The main purpose of this comparison study is to explore how the CR acceleration
depends on the preshock temperature and shock speed 
as well as shock Mach number.
Unlike pure gasdynamic shocks, CR modified shocks have intrinsic
scales, that is, a diffusion time,
$t_{\rm o}=\kappa_{\rm o}/u_{\rm o}^2$, and a diffusion length,
$x_{\rm o}=\kappa_{\rm o}/u_{\rm o}$,
because the mean acceleration time scale is $t_{acc} \propto \kappa(p)/u_{\rm o}^2$.
So the CR acceleration and the evolution of CR modified
shocks proceed faster for models with higher shock speeds. 
On the other hand, {\it two models with a same Mach number but with different
accretion speeds} should show qualitatively similar results, 
when the dynamical evolution is presented in terms of 
diffusion scales, $t_{\rm o}$ and $x_{\rm o}$,
since the CR acceleration is controlled mainly by the velocity
jump across the shock transition, which depends on shock Mach number.

For models with lower shock speed, the mean velocity of thermal particles 
is smaller and so the effective injection momentum, $p_{\rm inj}$, is smaller, 
compared with those for higher shock speed models.
Nonrelativistic suprathermal particles ($p_{\rm inj}<p<1$) are accelerated
quickly to relativistic energies because of small diffusion coefficients
($\kappa(p) \propto p^2$).
On the other hand, the highest momentum of the CR distribution 
increases with the age of the shock as $p_{\rm max} \sim 4.5 (t/t_{\rm o})$,
and so the CR distribution function extends to a similar value 
of $p_{\rm max}$ at a given $t/t_{\rm o}$ for all models.
As a result, for the model with lower shock speed, 
the ratio of $P_c/\rho_{\rm o} u_{\rm o}^2$ increases
faster, nonlinear modification to the underlying flow sets in earlier,
and the CR energy ratio is larger. 
However, we expect these differences would become smaller for $t/t_{\rm o}\gg1$
when relativistic particles dominate the CR pressure.
For example,
for the two models with the same Mach number at $M_s=6.8$ and the
same injection parameter at $\epsilon=0.2$,
$\Phi=0.32$ for $u_{\rm o}=1500\kms$ and $\Phi=0.39$ for $u_{\rm o}=75\kms$
at the terminal time. 
For larger values of $\epsilon$, $\Phi$ are higher 
due to the higher injection rate.  
For example, $\Phi=0.44$ for the model with $M_s=6.8$, 
$u_{\rm o}=75\kms$, and $\epsilon=0.3$.
Thus, for models with $M_s=6.8$, the CR energy ratio ranges 
$0.32\le \Phi \le 0.44$,  
depending on the shock speed and the injection parameter.
The difference becomes smaller at higher shock Mach numbers.
The CR energy ratio at the terminal time of our simulations is 
summarized for different models in Figure 7.

\acknowledgments{
This work was supported by KOSEF through Astrophysical Research Center
for the Structure and Evolution of Cosmos (ARCSEC).
The author would like to thank T.~W. Jones for helpful
comments on the manuscript.
}

\end{document}